# Neutron size from high energy scattering data

Michael J. Longo[1]

Department of Physics, University of Michigan, Ann Arbor, MI 48109, USA

**Abstract**

Alvarado, Aranda, and Bonilla propose a sub-GeV $U(1)_R$ gauge boson model to explain the proton charge radius discrepancy. Their model assumes opposite sign $U(1)_R$ charge assignments for up and down quarks in the nucleon. I point out that might cause a difference between neutron and proton mass radii that contradicts existing $p$-$p$ and $n$-$p$ scattering data. Using existing neutron and proton scattering data I show that the neutron and proton mass radii are equal within about 2%. This constitutes a challenge to models that attempt to explain the discrepancy by modifying the quark couplings.

There have been many studies of the proton charge radius using various methods. (For a recent review see the article by Gao and Vanderhaeghen[1].) A discrepancy $\approx 7\sigma$ persists between measurements with different methods [1]. This anomaly is still unresolved. The possibility of systematic effects has been much discussed. [1] Various "new physics" possibilities have been proposed to explain it, *e. g.,* [2, 3]. Some of these would affect the neutron structure differently than that of the proton as discussed below.

The internal charge distributions of the proton are much different. The charge radii of the proton and neutron are determined by their internal charge distributions. That in turn is determined by the strong interactions of the constituent quarks and gluons. Therefore, the charge radius of the proton is largely determined by strong interactions. Any model that attempts to explain the proton charge radius discrepancy must therefore be consistent with neutron and proton scattering data. Alvarado, Aranda, and Bonilla [2] propose an extension of the Standard Model with a $U(1)_R$ gauge symmetry in which only right-handed chiral fermions carry a nontrivial charge. In their model right-handed up and down quarks couple with opposite $U(1)_R$ charge assignments. Only the right-handed up quarks couple to the muon. The proton nominally is composed of 2 up quarks and one down quark, while the neutron has one up and 2 down. This could cause a significant effect on the neutron mass distribution compared to that of the proton.

---

[1] email: mlongo@umich.edu

The mass distributions of the neutron and proton can be compared via *n-p* and *p-p* total cross sections and elastic scattering cross sections. Figure 1, copied from Murthy *et al.* [4, 5], compares *n-p* and *p-p* total cross sections at GeV energies. We see from the fits that the cross sections agree to about 1% over the whole momentum range. In a simple gray disk optical model, the inelastic cross section is related to the radius as $\sigma_{inel} \approx \pi R^2 \xi$ where $\xi$ is the opacity, which is $\leq 1.0$. (See, for example, Ref. 6.) For a black disk, the elastic and inelastic cross sections are equal, so the total cross section is $\approx 2\pi R^2 \xi$. While the neutron and proton mass distributions are not at all like a black disk, a total cross section of 42 mb suggests an effective mass radius for both the *n-p* and *p-p* systems ~0.82 fm, which is comparable to the proton charge radius. Though these are combined radii for the *n-p* and *p-p* systems, we can still conclude that the neutron and proton mass radii are nearly equal.

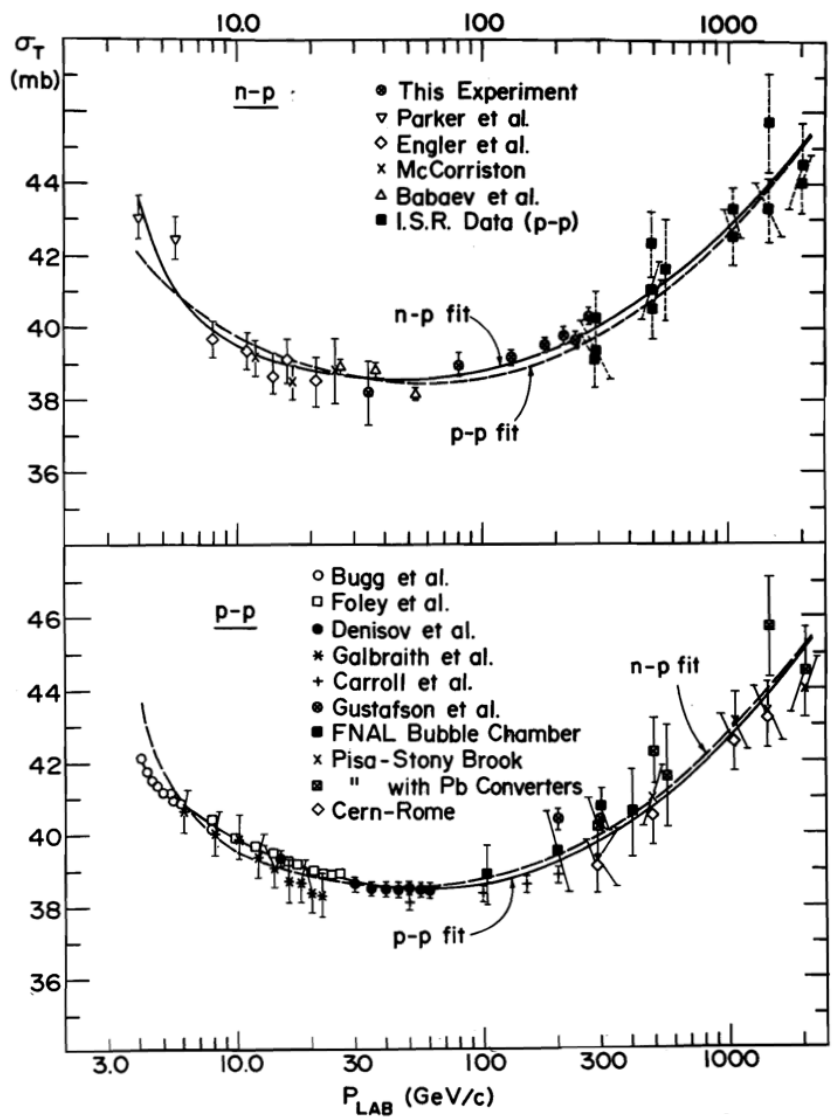

FIG. 1 – Total cross sections vs. lab momentum for *n-p* and *p-p* scattering from Murthy *et al.*[4]

Another measure of the neutron mass radius comes from *n-p* and *p-p* elastic scattering data at GeV energies. Figure 2 shows the logarithmic slope parameter for *n-p* and *p-p* elastic scattering from DeHaven *et al.* [7] plotted vs. *s*, the total energy in the center of mass system squared. These are evaluated at a four-momentum squared |*t*| of 0.2, which is well outside of the Coulomb region so the slope is determined by the mass distribution. The slope parameter is related to the nucleon form factors and mass radius. (See, for example, Ref. 8.) The slope parameters for *n-p* and *p-p* agree within ~3% up to the highest cms energies for which *n-p* data are available, ~700 GeV$^2$.

From the *n-p* and *p-p* data comparisons above we can conclude that the neutron and proton mass distributions are very similar at the 1–3% level, even though their charge distributions are totally different. Any model that tries to explain the proton charge radius discrepancy that involves the quark-gluon structure must take this into account.

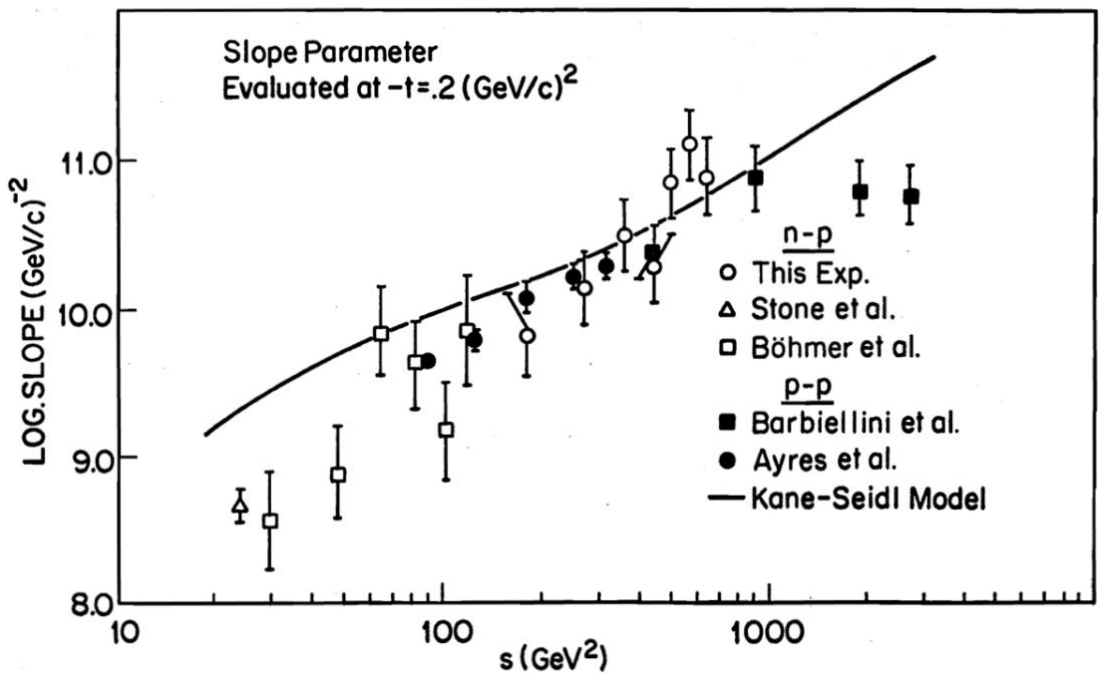

FIG. 2 – Logarithmic slope parameters vs. total cms energy squared for *n-p* and *p-p* scattering from DeHaven *et al.* [7]